\documentclass[aps,prl,twocolumn,floatfix,amsmath,amssymb,amsfonts,citeautoscript,longbibliography]{revtex4-2}
\usepackage[colorlinks=true,citecolor=blue]{hyperref}
\usepackage{graphicx}
\usepackage[percent]{overpic}
\usepackage{nicefrac}

\newcommand{\revision}[1]{{{#1}}}
\newcommand{\revisiontwo}[1]{{{#1}}}

\begin{document}
\title{Heat Pulses in Electron Quantum Optics}
\author{Pedro Portugal}
\affiliation{Department of Applied Physics, Aalto University, 00076 Aalto, Finland}
\author{Fredrik Brange}
\affiliation{Department of Applied Physics, Aalto University, 00076 Aalto, Finland}
\author{Christian Flindt}
\affiliation{Department of Applied Physics, Aalto University, 00076 Aalto, Finland}

\begin{abstract}
Electron quantum optics aims to realize ideas from the quantum theory of light with the role of photons being played by charge pulses in electronic conductors. Experimentally, the charge pulses are excited by time-dependent voltages, however, one could also generate heat pulses by heating and cooling an electrode. Here, we explore this intriguing idea by formulating a Floquet scattering theory of heat pulses in mesoscopic conductors. The adiabatic emission of heat pulses leads to a heat current that in linear response is given by the thermal conductance quantum. However, we also find a high-frequency component, which ensures that the fluctuation-dissipation theorem for heat currents, whose validity has been debated, is fulfilled. The heat pulses are uncharged, and we probe their electron-hole content by evaluating the partition noise in the outputs of a quantum point contact. We also employ a Hong--Ou--Mandel setup to examine if the pulses bunch or antibunch. Finally, to generate an electric current, we use a Mach--Zehnder interferometer that breaks the electron-hole symmetry and thereby enables a thermoelectric effect. Our work paves the way for systematic investigations of heat pulses in mesoscopic conductors, and it may stimulate future experiments.
\end{abstract}

\maketitle

\emph{Introduction.}--- Electron quantum optics is an emerging field of mesoscopic physics in which charge pulses are emitted into electronic circuits to realize interference experiments with electrons~\cite{Bocquillon:2014,Splettstoesser:2017,Bauerle:2018}. Clean single-particle excitations can be generated by applying Lorentzian voltage pulses to an electrode~\cite{Levitov:1996,Levitov:1997,Keeling:2006,Dubois:2013,Jullien:2014,Assouline:2023}, and the edge states of a quantum Hall sample can be shaped to form electronic circuits such as Mach--Zehnder~\cite{Ji:2003} or Fabry--P\'{e}rot interferometers~\cite{Ofek:2010}. Charges emitted close to the Fermi level can often be treated as non-interacting, for example, when interfering at an electronic beam splitter~\cite{Bocquillon:2013,Dubois:2013,Jullien:2014,Assouline:2023}. By contrast, at high energies, recent collision experiments have revealed strong interactions and non-linear effects~\cite{Wang:2023,Fletcher:2023,Ubbelohde:2023}. These advances show how experiments with time-dependent voltages bring about many exciting opportunities compared to their static counterparts.

Instead of generating charge pulses with a voltage, one could also excite heat pulses by heating and cooling an electrode to generate a time-dependent temperature~\cite{Portugal:2021,Portugal:2022,Chen:2023}. The generation of heat pulses in electron quantum optics is largely unexplored, and it could open up for a wide range of future applications and fundamental investigations~\cite{Dhar:2008,Battista:2013,Ludovico:2014,Pekola:2015,Dashti:2018,Pekola:2021,Majidi:2024}. For example, heat pulses should be electrically neutral, which might make them less prone to interaction and decoherence effects~\cite{Ferraro:2014}. Fast control of energy is also becoming increasingly important for heat management at the nano-scale and for converting waste heat into useful work~\cite{Samuelsson:2017,Li2023}. \revision{In addition, the accurate and precise control of heat pulses may find applications in many quantum technologies, including quantum computing and quantum sensing~\cite{Majidi:2024}.} Experiments with constant temperature biases have already observed the quantization of the thermal conductance in electronic conductors~\cite{Molenkamp:1992,Chiatti:2006,Jezouin:2013}, while \revision{topological} states have been shown to exhibit fractional thermal conductance~\cite{Banerjee:2018,Melcer:2023}. By contrast, much less is known about the propagation of heat pulses in mesoscopic conductors.

\begin{figure}[b!]
\begin{centering}
\includegraphics[width=0.98\columnwidth]{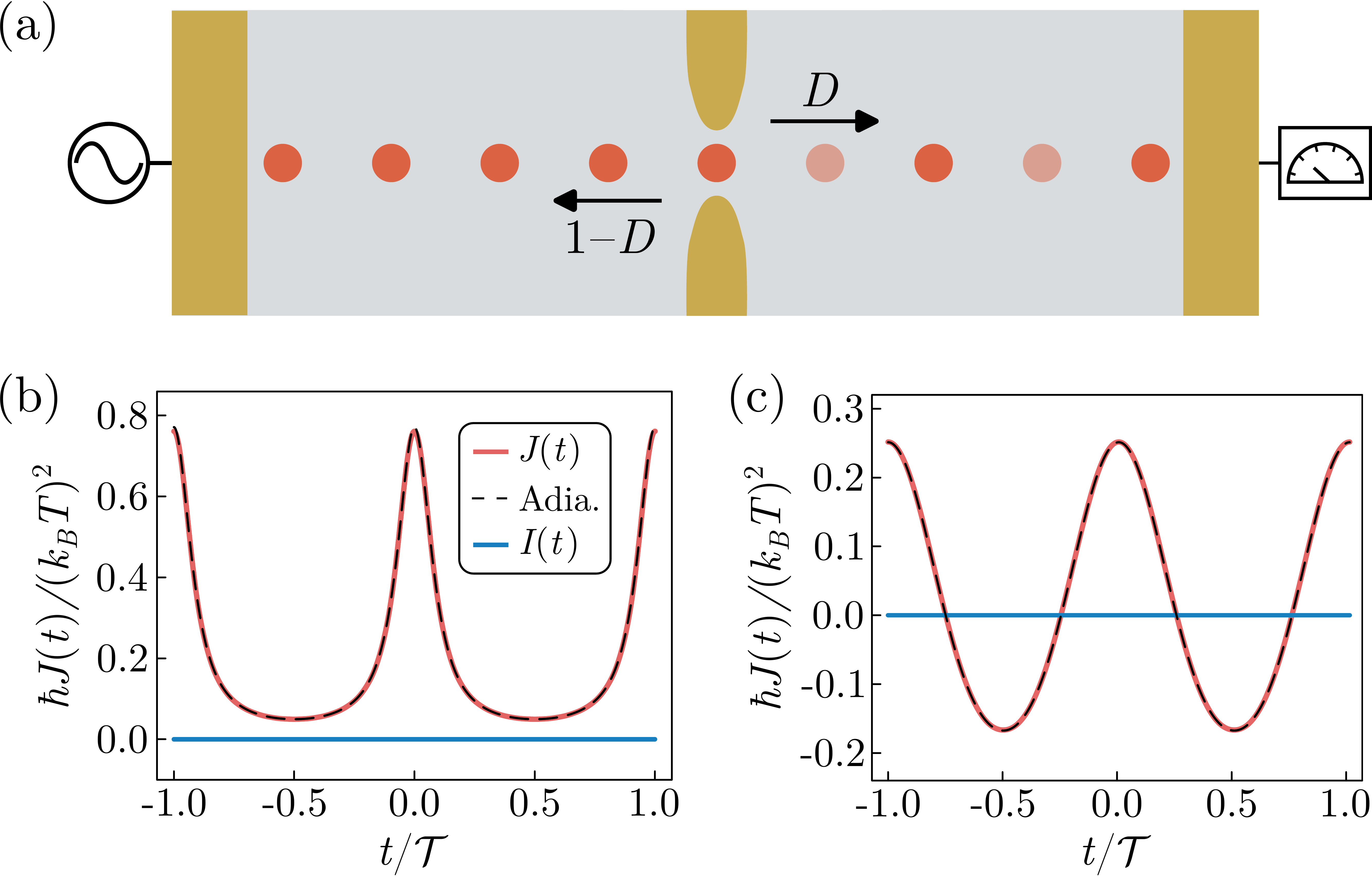}
\caption{Heat pulses in a mesoscopic conductor. (a)~\revision{A quantum point contact (central constriction in brown) with transmission $D$ is connected to source and drain electrodes (also brown). Heat pulses (red) are emitted from the source and propagate in a two-dimensional electron gas (gray) to the drain, where the electric current and the heat current are measured.} (b) Heat current for Lorentzian pulses of period~$\mathcal T$, half-width $\Gamma=0.1\mathcal
 T$, and amplitude $W_0=0.3$. The base temperature is $k_B T = 3\hbar\Omega$ with $\Omega=2\pi/\mathcal T$, and the quantum point contact is fully open, $D=1$. The dashed line indicates the static result in Eq.~(\ref{eq:Jstatic}) with the time-dependent temperature of the source electrode inserted, $T_S = [1+W(t)]T$, with $W(t)$ defined in Eq.~(\ref{eq:lorentzianpulse}). The electric current vanishes. (c) Similar results for a cosine drive with amplitude $W_0=0.4$. }
	\label{Fig1}
\end{centering}
\end{figure}

In this Letter, we formulate a Floquet scattering theory of heat pulses in mesoscopic conductors. To this end, we perform work on an electrode, which causes its temperature to change and leads to the periodic emission of heat pulses, see Fig.~\ref{Fig1}(a). \revision{We focus on the mesoscopic regime, where the base temperature is low enough (typically well below 100 mK) that the particle transport is coherent, and spurious heat leakage, for example due to phonons, can safely be neglected~\cite{Dubois:2013,Jullien:2014,Jezouin:2013,Assouline:2023,Majidi:2024}.  Our theory is valid for all realistic driving frequencies and driving amplitudes}. At low frequencies compared to the base temperature, the thermal conductance of a ballistic conductor is given by the thermal conductance quantum~\cite{Pendry:1983,Pekola:2021}. However, we also find a high-frequency component, which ensures that the fluctuation-dissipation theorem for heat currents, whose validity has been debated~\cite{Averin:2010,Zhan:2011,Lim:2013,Moskalets:2014}, indeed is fulfilled. We probe the number of electrons and holes in the uncharged heat pulses by evaluating the partition noise in the outputs of a quantum point contact~\cite{Dubois:2013,Dubois:2013b}. Moreover, to determine if the heat pulses bunch or antibunch, we consider a Hong--Ou--Mandel setup, where heat pulses are emitted onto each side of a quantum point contact~\cite{Hong:1987,Bocquillon:2013,Dubois:2013,Jullien:2014}. Finally, to generate an electric current, we employ a Mach–Zehnder interferometer that breaks the electron-hole symmetry and thereby enables a thermoelectric effect~\cite{Moskalets:2014b,Gaury:2014,Ryu:2022}. We focus here on mesoscopic conductors, but our findings may also be realized with atoms in optical lattices~\cite{Krinner:2014}.

\emph{Floquet scattering theory.}--- 
We consider the setup in Fig.~\ref{Fig1}(a), where a source and a drain electrode are connected by a quantum point contact. The central object of Floquet scattering theory is the scattering matrix, $S_n(E)$, which contains the amplitudes for an electron to be transmitted from the source to the drain after having changed its energy from $E$ to $E_n=E+n\hbar\Omega$ by exchanging $n$ modulation quanta with the external drive~\cite{Moskalets_book,Brandner:2020}. Here, the frequency $\Omega=2\pi/\mathcal{T}$ is given by the period of the drive $\mathcal{T}$. The heat pulses are generated far from the scatterer, such that the scattering matrix factorizes as $S_{n}(E)= S(E_n) K_n(E)$, where $S(E_n)$ describes the central scatterer, which in Fig.~\ref{Fig1}(a) is the quantum point contact, and $K_n(E)$ is the scattering matrix of the heat pulses, which we specify below. The electric current and the heat current are then measured in the drain electrode.

To generate heat pulses, we perform work on the source electrode as described by the Hamiltonian $\hat H(t)=\hat H_0+W(t)\hat H_0$, where $W(t)$ is the strength of the driving, and $\hat H_0$ is the Hamiltonian of the electrode without the drive. The chemical potential is set to zero, and energies are measured with respect to the Fermi level. Physically, the time-dependent Hamiltonian describes a process that heats or cools the electrode by performing work on it~\cite{aNote2}. Indeed, given the initial density matrix of the electrode at temperature $T_0$, $\hat\rho_0 = e^{-\hat H_0/k_B T_0}/Z_0$ with $Z_0=\mathrm{tr}\{e^{-\hat H_0/k_B T_0}\}$, one sees that it does not evolve in time, such that $\hat\rho(t) = \hat\rho_0$ at later times~\cite{bNoteSM}. However, we can rewrite it as $\hat \rho(t) =e^{-\hat H(t)/[1+W(t)]k_B T_0}/Z_0$, showing that the temperature changes as $T(t) = [1+W(t)]T_0$. We note that this approach resembles Luttinger's description of temperature~\cite{Luttinger:1964,Eich:2014,Tatara:2015,Bhandari:2020,Arrachea:2023,lopez:2023}. Also, it can be compared to the generation of charge pulses, where a time-dependent voltage is added to the Hamiltonian as $\hat H(t)=\hat H_0+eV(t)$. In that case, the voltage couples to charge, and charge pulses are emitted. By contrast, in our case, the driving field couples to energy, and heat pulses are emitted.

As an important result of this Letter, we find that the Floquet scattering matrix of the heat pulses reads
\begin{equation}
K(t,E)=\sqrt{1+W(t)}e^{-i E \phi(t)/\hbar},
\label{eq:floquetmatrix}
\end{equation}
where $\phi(t)=\int_{-\infty}^{t}dt'W(t')$ is the time-integrated driving field~\cite{bNoteSM}. If the drive has a constant offset, we decompose it as $W(t)=W_{\mathrm{dc}}+W_{\mathrm{ac}}(t)$ with $\int_{0}^{\mathcal{T}}dt W_{\mathrm{ac}}(t) =0$, such that the temperature reads $T(t) = [1+W_{\mathrm{dc}}][1+W_{\mathrm{ac}}(t)/(1+W_{\mathrm{dc}})]T_0$. The constant term in the first bracket is then included in the Fermi function of the source electrode, while we evaluate Eq.~(\ref{eq:floquetmatrix}) for the time-dependent part in the second bracket. We also define $K_n(E)=\int d t K(t,E)e^{i n \Omega t}/\mathcal T$ in the Fourier domain.
 
It is instructive to compare the scattering matrix with that of charge pulses, which only depends on the number of modulation quanta~$n$ and not on the initial energy~$E$~\cite{Note1}. In addition, the square-root prefactor in Eq.~(\ref{eq:floquetmatrix}) ensures a constant coupling to the central scatterer by compensating for the changed density of states in the source electrode~\cite{Arrachea:2006,Kato2018,Portugal:2021}. It is important to check that the scattering matrix is unitary, and after a few lines of algebra we indeed find $\sum_n K^*_{n-m}(E_m)	K_n(E) = \delta_{m,0}$ and $\sum_n K_{m-n}(E_n)K^*_{-n}(E_n) = \delta_{m,0}$, where the right-hand sides are one for $m=0$ and zero otherwise~\cite{bNoteSM}.

With the Floquet scattering matrix at hand, we can calculate the electric current in the drain electrode as~\cite{Moskalets_book}
\begin{equation}
		\! \! I(t) = \frac{e}{h}\int\limits_{-\infty}^{\infty}\! dE \sum_{n,l}\mathcal{F}_n(E) S^*_n(E)S_{n+l}(E)e^{-i\Omega t l},
		\label{eq:electriccurrent}
\end{equation}
while the heat current reads~\cite{Moskalets:2004,Moskalets_2016}
\begin{equation}
		\! \! J(t)\! = \!\! \! \int\limits_{-\infty}^{\infty}\! \! \frac{dE}{h} \sum_{n,l}\! E_{n+\frac{l}{2}}\mathcal{F}_{n+\frac{l}{2}}(E) S_n^*(E)S_{n+ l}(E)e^{-\scriptstyle{i\Omega t l}},
		\label{eq:heatcurrent}
\end{equation}
where $\mathcal{F}_n(E)=f_S(E)-f_D(E_n)$ is the difference between the Fermi functions of the source and the drain. More involved expressions exist for the noise spectra of the electric current~\cite{Moskalets_2002} and the heat current~\cite{Moskalets:2014}, and extensions to multi-terminal systems are also possible~\cite{Moskalets_book}.

\emph{Ballistic conductor.}--- We now consider heat that is emitted from the source electrode by periodic Lorentzians
\begin{equation}\label{eq:lorentzianpulse}
W(t) = \frac{W_0}{\pi}\sum_{n=-\infty}^\infty \frac{\Gamma \mathcal T}{\Gamma^2+(t-n\mathcal T)^2}
 \end{equation}
of half-width $\Gamma$ and amplitude set by $W_0$, or by a cosine drive reading $W(t) = W_0\cos{(\Omega t)}$. \revision{Experimentally, it has been found that voltage pulses that are produced by a pulse generator get deformed as they travel through the external circuit before reaching the mesoscopic device~\cite{Assouline:2023}. However, it is possible to characterize this deformation and compensate for it, when generating the pulses.} In Fig.~\ref{Fig1}(b,c), we show the heat current for slow drives, $\hbar\Omega\ll k_BT$, with the quantum point contact being fully open. We also show the electric current, which vanishes at all times. This conclusion can also be reached by examining Eq.~(\ref{eq:electriccurrent}) using the electron-hole symmetry $K_n(E)=K^*_{-n}(-E)$. For slow drives, the heat current should be given by the static expression~\cite{Pekola:2021}
\begin{equation}
	J = \frac{\pi^2k_B^2}{6h}(T_S^2-T_D^2)
	\label{eq:Jstatic}
\end{equation}
with the temperature of the source electrode replaced by the time-dependent one, $T_S(t)=[1+W(t)]T$, and the figure indeed confirms this expectation for both drives.

\begin{figure}
 \centering  
\includegraphics[width= 0.85\columnwidth]{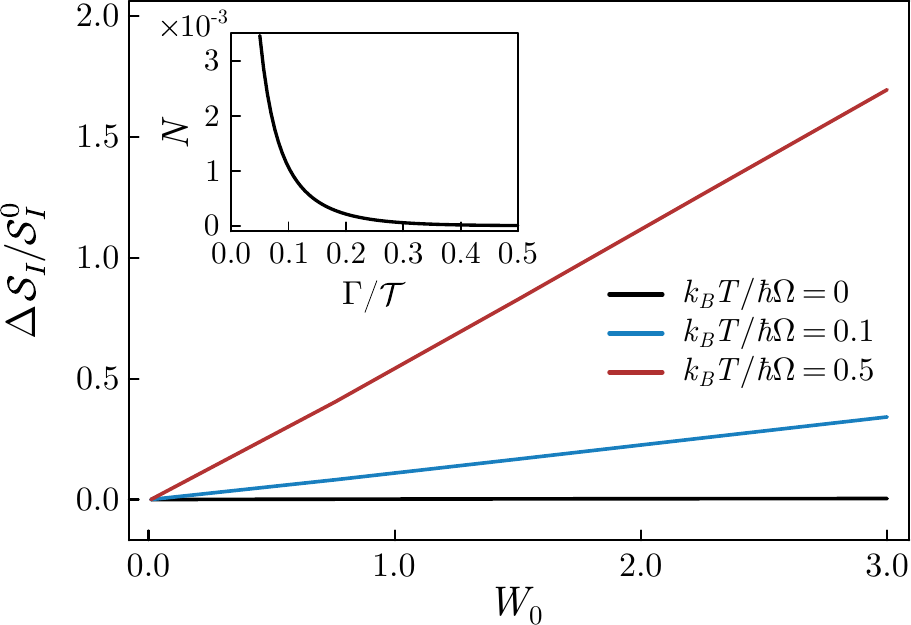}
\caption{Partition noise and number of charges in a heat pulse. Electric noise is generated by heat pulses that are partitioned on the quantum point contact with $D=1/2$. We show the difference between the noise with and without the drive, $\Delta\mathcal{S}_I = \mathcal{S}_I^{\mathrm{on}}(0) - \mathcal{S}_I^{\mathrm{off}}(0)$, normalized by the noise of a single particle $\mathcal{S}^{0}_I=e^2f D(1-D)$, where $f$ is the driving frequency. The half-width of the Lorentzian pulses is $\Gamma=0.2\mathcal T$. At zero temperature, this ratio yields the number of electrons and holes in a pulse as shown in the inset for~$W_0 = 0.2$.}
\label{Fig2}
\end{figure}

\emph{Linear response.}--- \revision{Our theory is valid for arbitrary driving strengths. However, it is instructive to} analyse the heat current in linear response to a \revision{weak} cosine drive with $W_0\ll 1$. To first order in $W_0$, we find $J(t)=\mathcal G(\Omega) W(t)$ with the frequency-dependent conductance
\begin{equation}
  \mathcal G(\omega)= \frac{(2\pi k_B T)^2+(\hbar\omega)^2}{12 h}.
  \label{eq:ther_cond}
\end{equation}
\revision{Using this expression for the conductance,} we can write the heat current as $J(t)=G_Q\Delta T(t)+ (f\hbar\Omega /12)W(t)$, where $G_Q= \pi^2 k_B^2 T/3h$ is the thermal conductance quantum, and $\Delta T(t)=W(t) T$ is the temperature bias. At high temperatures, the heat current is mainly driven by the temperature bias. On the other hand, as the temperature is lowered, the second contribution becomes relevant, and at zero temperature, the heat pulses are generated directly by the driving field, since the thermal conductance quantum vanishes. The magnitude of the heat current is then given by the size of the modulation quanta, $\hbar\Omega$, times the frequency at which they are emitted, $f=1/\mathcal T$. Importantly, the second term in Eq.~(\ref{eq:ther_cond}) ensures that the fluctuation-dissipation theorem for heat currents, whose validity has been debated~\cite{Averin:2010,Zhan:2011,Lim:2013,Moskalets:2014}, is fulfilled.  Indeed, by evaluating the equilibrium noise of the heat current~\cite{Sergi:2011}, we find $\mathcal{S}_Q(\omega)= \hbar\omega \mathrm{Re} [\mathcal G(\omega)]\coth{(\hbar\omega/2k_BT)}$ in accordance with the fluctuation-dissipation theorem~\cite{Kubo:1957}. \revisiontwo{(If the  driving field is not included in the Hamiltonian, one may find a thermal conductance that is frequency-independent, making it seem as if the fluctuation-dissipation theorem is violated.)} It is instructive to compare Eq.~(\ref{eq:ther_cond}) with the electric conductance of a ballistic conductor, which reads $e^2/h$ at all frequencies. 
Finally, we note that experiments with fast voltage pulses were conducted at temperatures below $T=50$ mK, corresponding to an energy of about $k_BT=0.4$ $\mu$eV, while the driving frequencies reached almost $\Omega= 20$ GHz, corresponding to the energy $\hbar\Omega =80$ $\mu$eV~\cite{Dubois:2013,Jullien:2014}. At such frequencies and temperatures, one would observe nonadiabatic effects according to Eq.~(\ref{eq:ther_cond}).

\begin{figure}
	 \centering  
	\includegraphics[width = 0.85\columnwidth]{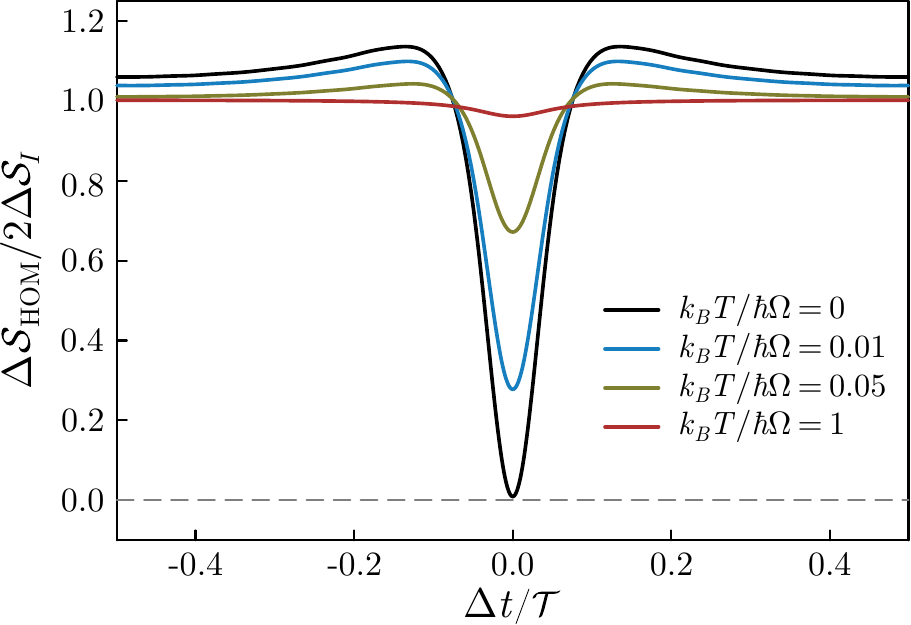}
	\caption{Hong-Ou-Mandel interferometry with heat pulses. Heat pulses arrive on each side of the quantum point contact with transmission $D=1/2$. We show the electric noise in the outputs as a function of a controllable time delay between the pulses. The parameters of the Lorentzian pulses are $W_0=0.1$ and $\Gamma=0.03\mathcal{T}$. At low temperatures, the noise is suppressed at zero time delay, showing that the heat pulses antibunch.}
	\label{Fig3}
\end{figure}

\emph{Partition noise.}--- The absence of an electric current in a ballistic conductor demonstrates that the heat pulses are electrically neutral. To probe the number of electrons and holes in a heat pulse, we proceed as in recent experiments and evaluate the electric noise in the outputs of the quantum point contact with a finite transmission~\cite{Dubois:2013,Dubois:2013b}. While electrons and holes contribute with opposite signs to the electric current, they generate the same partition noise, when scattered on a quantum point contact. Specifically, at zero temperature, the low-frequency noise can be expressed as $\mathcal{S}_I(0)=N\mathcal{S}^{0}_I$ in terms of the number of charges in a heat pulse, $N=N_e+N_h$, where $\mathcal{S}^{0}_I=e^2f D(1-D)$ is the partition noise of a single particle. Thus, from the low-frequency noise in the outputs, we can extract the number of charges as $N=\mathcal{S}_I(0)/\mathcal{S}^{0}_I$. At finite temperatures, we define the excess noise as the difference between the noise with and without the drive, $\Delta\mathcal{S}_I = \mathcal{S}_I^{\mathrm{on}}(0) - \mathcal{S}_I^{\mathrm{off}}(0)$, as shown in Fig.~\ref{Fig2} as a function of the driving amplitude, while the inset shows the number of electrons and holes in a pulse at zero temperature. \revisiontwo{As the pulses get narrower, they contain more high-frequency components, and more electron-hole pairs are generated. As the temperature is increased, the noise is mostly dominated by thermal fluctuations.}

\emph{Hong--Ou--Mandel interferometry.}--- We now consider heat pulses that are emitted onto each side of the quantum point contact to realize a Hong-Ou-Mandel setup~\cite{Hong:1987,Bocquillon:2013,Dubois:2013,Jullien:2014,Ferraro:2018}. Specifically, we evaluate the electric noise in the outputs as a function of a phase difference between the two drives. Identical bosons that arrive simultaneously on each side of a beam splitter are expected to bunch and leave via the same output arm. Fermions, by contrast, tend to antibunch and exit via different output arms. While bunching leads to increased noise, antibunching reduces the noise. Figure~\ref{Fig3} shows the excess noise for heat pulses that arrive on each side of the quantum point contact with a time delay between them. In this case, we normalize the noise with respect to the noise that is generated by pulses that arrive on only  one side of the quantum point contact. When the pulses arrive simultaneously, we observe a clear suppression of the noise due to antibunching. The width of the suppression is given by the overlap of the pulses, and the suppression is gradually lifted as the temperature is increased. Small enhancements appear around the suppressed noise, which have also been predicted for Hong--Ou--Mandel experiments with electron-hole collisions~\cite{Jonckheere:2012,Rech:2016}.

\begin{figure}
 \centering  
 \includegraphics[width = 0.85\columnwidth]{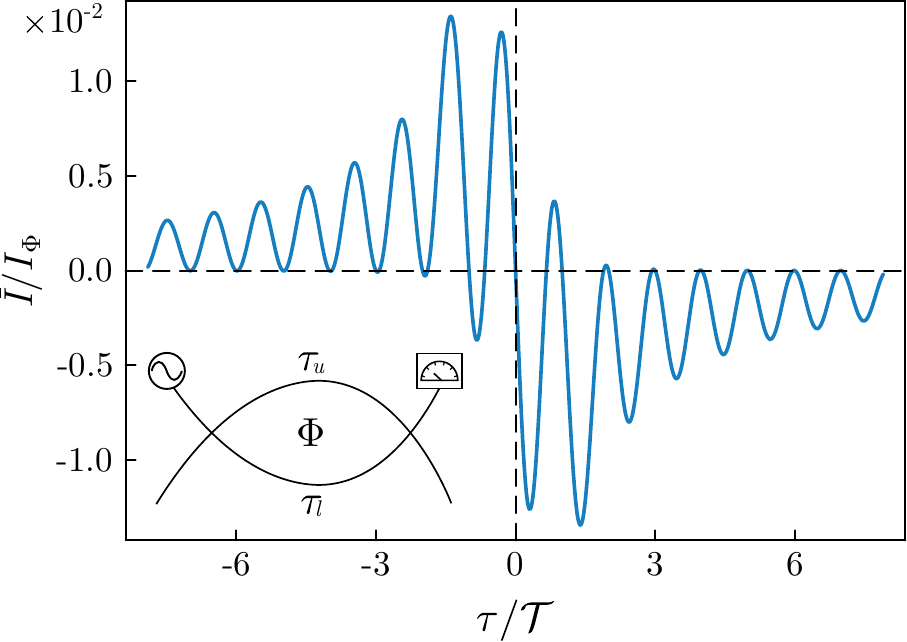}
  \caption{Mach--Zehnder interferometer and thermoelectric effect. Cosine pulses with amplitude $W_0$ are applied to the upper input of a Mach-Zenhder interferometer, and we show the average electric current in the upper output as a function of the path length difference to lowest order in the driving amplitude. The base temperature is zero. For non-zero fluxes and path-length differences, the electron-hole symmetry is broken, and a finite electric current is running in the upper output.}
  \label{Fig4}
\end{figure}

\emph{Mach--Zehnder interferometer.}--- The heat pulses are electrically neutral. However, we can generate an electric current using a Mach--Zehnder interferometer, which breaks the electron-hole symmetry and thereby enables a thermoelectric effect \cite{Moskalets:2014b}. We thus consider the interferometer in Fig.~\ref{Fig4} with the scattering amplitude
\begin{equation}
	S_{\mathrm{MZ}}(E) = r_1r_2e^{i(E\tau_u /\hbar+\phi_u)}+t_1t_2e^{i(E\tau_d/\hbar+\phi_d)}
	\label{eq:MZI}
\end{equation}
for transmissions from the upper input to the upper output. Here, the reflection (transmission) amplitudes of the first and second quantum point contact are denoted by $r_j$ ($t_j$), $j=1,2$, the time that it takes to transverse the upper and lower paths is $\tau_u$ and $\tau_d$, while $\phi_u$ and $\phi_d$ are the phases that are picked up due to an applied magnetic field. For half-transmitting quantum point contacts, $r_j,t_j=1/\sqrt{2}$, the transmission probability reads $T_{\mathrm{MZ}}(E)=|S_{\mathrm{MZ}}(E)|^2=[1+\cos(E\tau/\hbar+\Phi)]/2$, where $\tau=\tau_u-\tau_d$ is given by the path-length difference, and $\Phi=\phi_u-\phi_d$ is controlled by the enclosed magnetic flux. Importantly, the magnetic flux allows us to break the electron-hole symmetry, so that $T_{\mathrm{MZ}}(-E)\neq T_{\mathrm{MZ}}(E)$.

In Fig.~\ref{Fig4}, we show the average electric current in the upper output as a function of the path length difference. For a cosine drive, the electric current at zero temperature can be calculated to lowest order in the amplitude, 
\begin{equation}
	\bar{I}= I_\Phi \sin ^2( \bar\tau ) \left[1-\bar\tau \cot ( \bar\tau )- \bar\tau
  ^2/2\right]/(2\bar\tau)^3 ,
	\label{eq:MZIcurr}
\end{equation}
where we have defined $I_\Phi = W_0^2 e f \sin (\Phi )$ and $\bar \tau = \pi\tau/\mathcal{T}$. From this expression, we see that an electric current indeed can be generated if a magnetic field is applied, so that $I_\Phi\neq 0$, and the current can be either positive or negative depending on the flux. The current also changes sign as $\bar{I}(-\tau)=-\bar{I}(\tau)$, and it vanishes for path length differences that match the period of the drive, $\tau=n\mathcal T$. In addition, there is no current for large path length differences. These results show how the sign and the magnitude of the electric current can be controlled by the magnetic flux and the path length difference. \revisiontwo{One could also use a Fabry-Pérot interferometer~\cite{Gaury:2014} or a quantum dot~\cite{Ryu:2022} to generate a thermoelectric effect.}

\emph{Conclusions.}--- We have presented a Floquet scattering theory of heat pulses in mesoscopic conductors. The pulses are generated in response to work that is performed on an electrode to change its temperature. We have found the response function for a ballistic conductor, which agrees with the fluctuation-dissipation theorem for heat currents. The pulses are uncharged, and the number of electrons and holes in a pulse can be determined from the partition noise after a quantum point contact. We have also found that the heat pulses antibunch when arriving simultaneously on each side of a quantum point contact. Finally, we have used a Mach-Zehnder interferometer to break the electron-hole symmetry and thereby enable a thermoelectric effect.

Our work paves the way for systematic investigations of dynamic heat transport in mesoscopic conductors, and it can be extended in many different directions. For example, it would be interesting to visualize the heat pulses using their Wigner representation. The influence of the pulse shape on the emitted pulses should also be explored. Since the heat pulses are electrically neutral, it is possible that they are less prone to interactions and decoherence effects, which is another question that deserves to be addressed. Moreover, while we have mainly focused on the properties of the heat pulses themselves, it would be relevant to develop a microscopic description of how work is performed on an electrode to generate the pulses, for example by shining light on it. 
\revision{Finally, while we have focused on phase-coherent transport in mesoscopic conductors at low temperatures, it would be interesting to investigate dynamic heat transport in interacting nanostructures, such as quantum dots, or in situations, where heat may be lost to phonons in the host material. To this end, it would be useful to adapt the Meir-Wingreen formula to describe time-modulated temperatures~\cite{Jauho:1994}.}

\emph{Acknowledgements.}--- We~thank~K.~Brandner,~P.~Burset,~K.~Flensberg,~C.~Gorini,~N.~Lo~Gullo, M.~V.~Moskalets, B.~Roussel, and P.~Samuelsson for useful discussions. We acknowledge support from Jane and Aatos Erkko Foundation, the Research Council of Finland through the Finnish Centre of Excellence in Quantum Technology (352925), and the Japan Society for the Promotion of Science
through an Invitational Fellowship for Research in Japan.

\end{document}


\title{Supplemental Material for ``Heat Pulses in Electron Quantum Optics"}
\author{Pedro Portugal}
\affiliation{Department of Applied Physics, Aalto University, 00076 Aalto, Finland}
\author{Fredrik Brange}
\affiliation{Department of Applied Physics, Aalto University, 00076 Aalto, Finland}
\author{Christian Flindt}
\affiliation{Department of Applied Physics, Aalto University, 00076 Aalto, Finland}

\newcommand{\revision}[1]{{{#1}}}

\maketitle

\onecolumngrid
%
%
%
%
\section{Floquet scattering matrix}
\label{sec:floq_scat}
\noindent
We start by considering a particle reservoir, which is prepared in the equilibrium state $\hat\rho_0 = e^{-\hat H_0/k_B T_0}/Z_0$, where $\hat H_0$ is the Hamiltonian, the temperature is denoted by $T_0$, and $Z_0=\mathrm{tr}\{e^{-\hat H_0/k_B T_0}\}$ is the partition function. We have set the chemical potential to zero, which is not important in the following. We now perform work on the reservoir to change its temperature. To this end, we consider the time-dependent Hamiltonian $\hat H(t) = \hat H_0+ W(t)\hat H_0=[1+W(t)]\hat H_0$, where $W(t)$ is the amplitude of the driving field.  Since the Hamiltonian commutes with itself at different times and the initial density matrix, the density matrix is constant, such that $\hat \rho(t) = \hat\rho_0$ at all times. \revision{To see this, we note that the density matrix evolves according to the von Neumann equation,
	\begin{equation}
	\frac{d}{dt}\hat{\rho}(t)=\frac{1}{i\hbar}[\hat H (t), \hat{\rho}(t)]=\frac{1+W(t)}{i\hbar}[\hat H_0, \hat{\rho}(t)].
	\end{equation}	
Integrating this equation of motion once and iterating, we find  
	\begin{equation}
		\begin{split}
	\hat{\rho}(t)&=\hat{\rho}(0)+\int_0^t dt' \frac{1+W(t')}{i\hbar}[\hat H_0, \hat{\rho}(t')]\\
	&=\hat{\rho}(0)+\int_0^t dt' \frac{1+W(t')}{i\hbar}[\hat H_0, \hat{\rho}(0)]+\int_0^t \int_0^{t'} dt'dt'' \frac{1+W(t')}{i\hbar}\frac{1+W(t'')}{i\hbar}[\hat H_0, [\hat H_0, \hat{\rho}(0)]]+\ldots
	\end{split}
\end{equation}	
Since $\hat{\rho}(0)$ is given by the exponential of $\hat H_0$, it commutes with $\hat H_0$, so all commutators above vanish, and we are left with only the first term in the expression. Thus, we can conclude that $\hat{\rho}(t)=\hat{\rho}(0)$. However, although, the density matrix does not change over time,} we can rewrite it as $\hat \rho(t) =e^{-\hat H(t)/[1+W(t)]k_B T_0}/Z_0$, showing that the time-dependent temperature reads $T(t) = [1+W(t)]T_0$. Thus, by performing work on the reservoir, its temperature changes.

To proceed, we write the time-dependent Hamiltonian  as
\begin{equation}
    H(t)  = \left[1+W(t)\right]\int d E E \hat c_0^\dagger(E) \hat c_0(E),
\end{equation}
where the operator $\hat c_0^\dagger(E)$ creates the Floquet states $\Psi(E,t) = e^{-i E (t+\phi(t))/\hbar} \psi_0(E)$,
the eigenstates of $\hat H_0$ are denoted as $\psi_0(E)$, and  $\phi(t) = \int_{-\infty}^t dt^\prime W(t^\prime)$ is the accumulated phase. To find the scattering amplitudes, we note that the density of states is changed by the factor $1+W(t)$. To account for this change, the scattering amplitudes have to be renormalized by a factor of $\sqrt{1+W(t)}$. The wave function for  particles leaving the driven reservoir therefore reads
\begin{equation}
    \Tilde{\Psi}(E,t) =K(t,E) e^{-i E t/\hbar}\psi_0(E),
\end{equation}
where
\begin{equation}
K(E,t) = \sqrt{1+W(t)}e^{-i E \phi(t)/\hbar}
\label{eq:scat_mat}
\end{equation}
is the scattering matrix of the main text. We now assume that the drive has the period $\mathcal{T}$, such that $W(t+\mathcal{T}) = W(t)$.  We also write it as $W(t)=W_{\mathrm{dc}}+W_{\mathrm{ac}}(t)$, so that the time-dependent part integrates to zero, $\int_0^{\mathcal{T}} dt W_{\mathrm{ac}}(t)=0$, while the constant part can be absorbed into the constant temperature of the reservoir. Specifically, we can rewrite the temperature as $T(t)=[1+W(t)]T_0= [1+\tilde W_{\mathrm{ac}}(t)] \tilde T_0$ with $\tilde T_0=(1+W_{\mathrm{dc}})T_0$ and $\tilde W_{\mathrm{ac}}(t)=W_{\mathrm{ac}}(t)/(1+W_{\mathrm{dc}})$.

Next, we define the Fourier series
\begin{equation}
    K(t,E) = \sum_{n=-\infty}^\infty K_n(E) e^{-i n\Omega t/\hbar},
\end{equation}
where $\Omega = 2 \pi/\mathcal{T}$ is the frequency of the drive, and the Fourier components are
\begin{equation}
    K_n(E) = \int_0^{\mathcal T}\frac{d t}{\mathcal T} K(t,E) e^{i n\Omega t/\hbar}.
\end{equation}
We can then write
\begin{equation}
 \Tilde{\Psi}(E,t) = e^{-i E t/\hbar}\sum_{n=-\infty}^\infty  e^{-i n\Omega t}K_n(E)\psi_0(E)=\sum_{n=-\infty}^\infty  e^{-i E_n  t/\hbar}\tilde\psi_{0}(E_n),
 \end{equation}
where we have defined $E_n =E+n\hbar\Omega$ and $\tilde\psi_{0}(E_n)= K_n(E)\psi_0(E)$. We can thereby identify $K_n(E) $ as the scattering matrix that describes the interaction of a particle with the time-dependent drive, leading to a change of its  energy from $E$ to $E_n$. In second quantization, this relation can be expressed as
\begin{equation}
\hat c(E) = \sum_{n=-\infty}^\infty K_{n}(E_{-n}) c_0(E_{-n}),\hspace{0.5 cm}
\hat c^\dagger(E) = \sum_{n=-\infty}^\infty K^*_{n}(E_{-n}) c^\dagger_0(E_{-n}),
\end{equation}
where the annihilation and creation operators on the left-hand side describe particles that are emitted from the driven reservoir. We note that the factor  of $\sqrt{1+W(t)}$ in Eq.~(\ref{eq:scat_mat}) ensures that the spectral density is constant in accordance with a generalized Fisher-Lee approach for deriving scattering matrices~\cite{Portugal:2021,Arrachea:2006}.

%
%
%
%
\section{Unitarity of the Floquet scattering matrix}
\label{sec:unitarity}
\noindent
To demonstrate that the Floquet scattering matrix of the heat pulses is unitary, we use that
\begin{equation}
	\begin{split}
		\sum_n K^*_{n-m}(E_m)K_{n}(E)
		=& \int_0^\mathcal{T} \frac{dt}{\mathcal T} \int_0^\mathcal{T} \frac{dt^\prime}{\mathcal T} \left[\sum_n e^{in\Omega (t^\prime-t)}\right] e^{im\Omega t}  K^*(t,E_m) K(t^\prime,E)=\int_0^\mathcal{T} \frac{dt}{\mathcal T} e^{im\Omega t}  K^*(t,E_m) K(t,E) \\
		& = \int_0^\mathcal{T} \frac{dt}{\mathcal T}  [1+W(t)]e^{i m \Omega \phi(t)} = \int_{\phi(0)}^{\phi(\mathcal{T})}  \frac{d\phi}{\mathcal{T}} e^{i m \Omega \phi} = \delta_{m,0},
	\end{split}
\end{equation}
since $\sum_n e^{i n \Omega (t- t^\prime)}=\mathcal{T} \sum_n \delta[\mathcal{T} n-(t-t^\prime)]$ in the first line, and $\int_0^\mathcal{T}dt W(t) = 0$ in the second line. Similarly, we find
\begin{equation}
	\begin{split}
	\sum_n K_{m-n}(E_n)K_{-n}^*(E_n) =& \int_0^\mathcal{T} \frac{dt}{\mathcal T} \int_0^\mathcal{T} \frac{dt^\prime}{\mathcal T}  e^{im\Omega t}  \sqrt{1+W(t)}\sqrt{1+W(t^\prime)} e^{-i \int_{t^\prime}^{t} d s W(s) E/\hbar} \sum_n e^{-i n \Omega\int_{t^\prime}^{t} d s  [1+W(s)]}\\
		=& \int_0^\mathcal{T} \frac{dt}{\mathcal T} \int_0^\mathcal{T}dt^\prime e^{im\Omega t}  \sqrt{1+W(t)}\sqrt{1+W(t^\prime)} e^{-i \int_{t^\prime}^{t}d s W(s) E/\hbar} \sum_n \delta\left(\mathcal T n - \int_{t^\prime}^t d t^\prime [1+W(t^\prime)]\right)\\
		=& \int_0^\mathcal{T} \frac{dt}{\mathcal T}  e^{im\Omega t} = \delta_{m,0},
	\end{split}
\end{equation}
where we have used the explicit expression for $K(t,E)$ in Eq.~\eqref{eq:scat_mat}.

%
%
%
%
\section{Lorentzian and cosine pulses}
\noindent
To obtain the Floquet scattering matrix for specific shapes of the drive with $|W(t)|<1$, we expand it as
\begin{equation}
K_n(E)=\int_0^{\mathcal T}\frac{d t}{\mathcal T}
K(t,E)e^{i n \Omega t} =  \int_0^{\mathcal T}\frac{d t}{\mathcal T}\sqrt{1+W(t)} e^{-i E\phi(t) /\hbar}e^{i n \Omega t}=\int_0^{\mathcal T}\frac{d t}{\mathcal T}\sum_{m=0}^\infty \binom{1/2}{m} W^m(t) e^{-i E\phi(t) /\hbar}e^{i n \Omega t}
\label{eq:Floq-eval}
\end{equation}
in terms of the binomial coefficients. We take the phase $\phi(t)$ as the integral of $W(t)$ from $0$ to $t$, since the integral over earlier times has no implications. We can then evaluate the Floquet scattering matrices for different drives. 

We first consider a series of Lorentzian pulses of temporal width $\Gamma$ and period $\mathcal T$, which can be written as~\cite{Dubois_2013b} 
\begin{equation}
        W(t) = \frac{W_0}{\pi}\sum_{n=-\infty}^\infty \frac{\Gamma \mathcal T}{\Gamma^2+(t-n\mathcal T)^2}
        =\frac{W_0}{2}\frac{ \sinh(2 \pi \Gamma/\mathcal T)}{\sinh^2 (\pi \Gamma/\mathcal T)+\sin^2( 
 \Omega t/2)}.
 \end{equation}
We decompose the pulse as $W(t)=W_{\mathrm{dc}}+W_{\mathrm{ac}}(t)$  with $W_{\mathrm{dc}}=W_0$. Thus, in the following, we evaluate the Floquet scattering matrix for the  time-dependent part, $W_{\mathrm{ac}}(t)$, and we skip the subscript to simplify the notation. 
 
 The phase corresponding to the time-dependent part can be calculated analytically and reads~\cite{Dubois_2013b}
 \begin{equation}
    e^{-i E\phi(t)/\hbar} = \left(e^{i  \Omega t }\frac{\sin(\pi(t + i \Gamma)/\mathcal{T})}{\sin (\pi ( t - i \Gamma)/\mathcal{T})}\right)^{W_0 \omega}, 
\end{equation}
where we have defined $\omega = E/{\hbar\Omega}$.
By also defining $z=e^{i \Omega t}$ and $\gamma = e^{-2\pi \Gamma/\mathcal{T}}$, we can rewrite the pulse as
\begin{equation}
        W(t) = W_0\left(\frac{1}{1- \gamma z^*} + \frac{1}{1-\gamma  z}-2 \right).
\end{equation}
Inserting this expression into Eq.~(\ref{eq:Floq-eval}), we find
\begin{equation}
       K_n(E)=\sum_{m=0}^\infty  W_0^m \binom{\nicefrac{1}{2}}{m}\oint_{|z| = 1}\frac{ d z}{ 2 \pi i z}\left( \frac{1}{1- \gamma  z^* } + \frac{1}{1-\gamma  z}-2 \right)^m \left(\frac{1-\gamma  z}{1-\gamma  z^*}\right)^{W_0 \omega} z^n,
\end{equation}  
where we have used that
\begin{equation}
        \frac{d t}{\mathcal{T}}\left(e^{i \Omega t}\frac{\sin(\pi(t + i \Gamma)/\mathcal{T})}{\sin (\pi (t - i \Gamma)/\mathcal{T})}\right)^{W_0 \omega}e^{i n \Omega  t} =\frac{d z}{2 \pi i z} z^n \left(\frac{1-\gamma  z}{1-\gamma  z^*}\right)^{W_0 \omega}.
\end{equation}
We then finally arrive at the expression
\begin{equation}
       K_n(E)=(-1)^{n}\sum_{m=0}^\infty \binom{\nicefrac{1}{2}}{m} W_0^m \sum_{l=0}^{\infty}  \gamma^{2 n+l} \sum_{q=0}^n\sum_{p=0}^{n-q} (-2)^{m-q-p}\binom{m}{q,p,m-q-p}    \binom{W_0 \omega - p}{l}\binom{-W_0 \omega -q}{l+n} 
\end{equation}
where the multinomial expansion was used together with Cauchy's residue theorem and an expansion around $\gamma =0$.

For a cosine pulse of the form $W(t) = W_0 \cos(\Omega t)$, we make extensive use of Feynman's integral trick by noting that
\begin{equation}
          \sin^{n} (\Omega t)e^{-i  \chi W_0\omega\sin(\Omega t)}=(i/ W_0\omega  )^{n}\partial_\chi^{n}e^{-i  \chi W_0\omega\sin(\Omega t)}.
\end{equation}
We then find
\begin{equation}
\begin{split}
K_n(E)&=\int \frac{d t}{\mathcal T}\sum_{m=0}^\infty \binom{1/2}{2 m} W_0^{2 m}\cos^{2m}{(\Omega t)} (1+W_0\cos{(\Omega t)}) e^{-i W_0 \sin(\Omega t) \omega}e^{i \Omega t}\\
&=\int \frac{d t}{\mathcal T}\sum_{m=0}^\infty \binom{1/2}{2 m} W_0^{2 m} (1+W_0 \cos{(\Omega t)})(1-\sin^2{(\Omega t)})^{m} e^{-iW_0\sin(\Omega t)\omega}e^{i  \Omega t},
\end{split}
\end{equation}
and finally
\begin{equation}
\begin{split}
K_n(E)=&\sum_{m=0}^\infty\sum_{k=0}^m\sum_{q=0}^{2k} \binom{\nicefrac{1}{2}}{2m}\binom{m}{k}\binom{2k}{q}\frac{(-1)^{q}W_0^{2m}}{2^{2k}}\\
&\times\left[ J_{n+2q-2k}(
W_0 \omega) + \frac{W_0}{2}\frac{(1-4m)}{4m+2}\Big(J_{n+2q-2k+1}(
W_0 \omega)+J_{n+2q-2k-1}(
W_0 \omega)\Big)\right],
\end{split}
\end{equation}
having used the Jacobi-Anger expansion in terms of the Bessel functions of the first kind, 
\begin{equation}
e^{i z \cos \theta}=\sum_{n=-\infty}^\infty i^n J_n(z) e^{i n \theta}.
\end{equation}

%
%
%
%
\section{Electric current}
\label{sec:electricalcurrent} 
\noindent 
We now demonstrate that the heat pulses are electrically neutral. The electric current of a ballistic conductor reads
\begin{equation}
I(t) = \frac{e}{h} \sum_{n,m=-\infty }^\infty \int_{-\infty}^\infty d E e^{-im\Omega t } K^*_n(E) K_{n+m}(E) \left[ f_S(E)-f_D(E_n)\right ].
\end{equation}
We first identify the Fourier coefficients of the time-dependent current as
\begin{equation}
I_m = \frac{e}{h} \sum_{n=-\infty}^\infty \int_{-\infty}^\infty d E K^*_n(E) K_{n+m}(E) \left[ f_S(E)-f_D(E_n)\right ],
\end{equation}
such that $I(t) = \sum_{m=-\infty}^\infty I_m e^{-im\Omega t}$. Since the electric current is real, we have $I_m = I_{-m}^*$. We therefore investigate
\begin{equation}
    I_m+I_{-m}^* = \frac{e}{h}\sum_n \int dE \left(K_n^*(E) K_{n+m}(E)+K_n(E) K_{n-m}^*(E)\right)[f_S(E) - f_D(E_n)].
    \label{eq:symmetrized}
\end{equation}
We also have
\begin{equation}
K_n^*(E) = \int_{0}^{\mathcal T} \frac{dt}{\mathcal T} K(t,E)^* e^{-in\Omega t}=\int_{0}^{\mathcal T}\frac{dt}{\mathcal T} K(t,-E) e^{-in\Omega t}=K_{-n}(-E),
\end{equation}
which we insert into Eq.~\eqref{eq:symmetrized}, which becomes
\begin{equation}
    I_m+I_{-m}^* = \frac{e}{h}\sum_n \int dE \big(K_{-n}(-E) K_{n+m}(E)+K_n(E) K_{-n+m}(-E)\big)[f_S(E) - f_D(E_n)].
\end{equation}
In this expression, $K_{-n}(-E) K_{n+m}(E)+K_n(E) K_{-n+m}(-E)$ is invariant under a change of sign of both $n$ and $E$. Under this transformation, we also have
\begin{equation}
f_S(-E)-f_D(-E_{-n}) = f_S(-E) -f_D(-E-n\hbar \Omega) \\
= [1-f_S(E)]-[1-f_D(E+n\hbar \Omega)] = - \left[ f_S(E)-f_D(E_n)\right ],
\end{equation}
where we have used that $f(-E) = 1-f(E)$ for the Fermi functions. Thus, any pair $(n,E)$ with $n\neq0$ yields an integrand, which has a counterpart for $(-n,-E)$ with the same magnitude, but opposite sign. Thus, we find
\begin{equation}
    I_m+I_{-m}^* = \frac{e}{h}\int dE \big(K_{0}(-E) K_{m}(E)+K_0(E) K_{m}(-E)\big)[f_S(E) - f_D(E)] = 0 ,
\end{equation}
since there is no constant bias. We therefore have $I_m = 0$ as well as $I(t) = 0$.

%
%
%
%
\section{Linear-response theory}
\label{sec:linear response}
\noindent
The starting point for our linear-response theory is the expression for the heat current in a ballistic conductor~\cite{Moskalets_2016},
\begin{equation}
J(t) = \frac{1}{h}\sum_{m,n } \int d E E_{n+m/2} e^{-im\Omega t }K_{m+n}(E)K_n^*(E) \left[f(E)- f(E_{n+m/2})\right ],
\label{eqJ}
\end{equation}
where we take the same Fermi functions for both reservoirs, since we consider a small-amplitude drive of the form $W(t) = W_0 \cos(\Omega t)$ with no constant temperature bias. To first order in $W_0\ll 1$, we find
\begin{equation}
K_n(E) = \int_0^\mathcal{T} \frac{dt}{\mathcal{T}} \sqrt{1+W(t)}e^{-iE\int_{-\infty}^t dt^\prime W(t^\prime)/\hbar}e^{i n\Omega t}\simeq \int_0^\mathcal{T} \frac{dt}{\mathcal{T}} \left(1+\frac{W_0 \cos(\Omega t)}{2}-\frac{i W_0 E}{\hbar} \left[ \frac{\sin(\Omega t)}{\Omega}+c  \right ] \right) e^{i n\Omega t}, 
\end{equation}
where $c= \int_{-\infty}^0 dt' \cos(\Omega t')$ can be omitted, since it is an overall phase with no physical implications. We then find
\begin{equation}
K_n(E) \simeq\delta_{n,0}+W_0 \left(\frac{1}{4}-\frac{E}{2\hbar \Omega} \right)\delta_{n,-1}+W_0\left(\frac{1}{4}+\frac{E}{2\hbar \Omega} \right)\delta_{n,1}.
\end{equation}
Inserting this result into Eq.~\eqref{eqJ}, we obtain to first order in $W_0$ the expression
\begin{equation}
J(t)\simeq \frac{(2\pi k_B T)^2+(\hbar\Omega)^2}{24 \pi \hbar}W_0\cos(\Omega t) = \mathcal{G}(\Omega) W(t),
\end{equation}
with
\begin{equation}
    \mathcal G(\omega)= \frac{(2\pi k_B T)^2+(\hbar\omega)^2}{24 \pi \hbar}.
\label{eq:linG}
\end{equation}
Since the current is linear in the driving strength, the general result for an arbitrary pulse $W(t)$ becomes
\begin{equation}
    J(t) \simeq  G_Q W(t)  T - \frac{\hbar}{24\pi  }\partial_t^2 W(t),
\end{equation}
where $G_Q=\pi^2k_B^2 T/ 3h$ is the thermal conductance quantum, and we have used that
\begin{equation}
    \sum_n W_n \cos(n \Omega t) (n\Omega)^2= -\partial_t^2 W(t),
\end{equation}
where $W_n$ are the Fourier components of $W(t)$.

%
%
%
%
\section{Fluctuation-dissipation theorem}
\noindent
According to the fluctuation-dissipation theorem, the equilibrium noise of the heat current can be expressed as 
\begin{equation}
\mathcal{S}_Q(\omega)= \hbar\omega \mathrm{Re} [\mathcal G(\omega)]\coth{(\hbar\omega/2k_BT)}
\label{eq:FDT}
\end{equation}
in terms of the conductance $\mathcal G(\omega)$ in linear response to the field $W(t)$ that drives the heat current~\cite{Kubo:1957}. To confirm this relation, we calculate the equilibrium noise of the heat current for a ballistic conductor, which reads~\cite{Sergi:2011}
\begin{equation}
\mathcal{S}_Q(\omega)= \frac{1}{h}\int_{-\infty}^{\infty}dE(E+\hbar\omega)^2[f(E)(1-f(E+\hbar\omega))+f(E+\hbar\omega)(1-f(E))],
\label{eq:equi_noise}
\end{equation}
where $f(E)$ is the Fermi function. We can rewrite this expression as
\begin{equation}
\mathcal{S}_Q(\omega)= \frac{n_B(\hbar\omega)+n_B(-\hbar\omega)}{h}\int_{-\infty}^{\infty}dE(E+\hbar\omega)^2[f(E)-f(E+\hbar\omega)],
\end{equation}
where $n_B(E)$ is the Bose-Einstein distribution. Now, carrying out the integral, we find
\begin{equation}
\mathcal{S}_Q(\omega)= \frac{(2\pi k_B T)^2+(\hbar\omega)^2}{24 \pi \hbar }\hbar\omega \coth{(\hbar\omega/2k_BT)}= \hbar\omega \mathrm{Re} [\mathcal G(\omega)]\coth{(\hbar\omega/2k_BT)},
\end{equation}
where we have used Eq.~(\ref{eq:linG}) for the linear response function. Thus, we see that the fluctuation-dissipation theorem for heat currents indeed is fulfilled.

\clearpage
\newpage